\documentclass[fleqn,usenatbib]{mnras}

\usepackage{newtxtext,newtxmath}

\usepackage[T1]{fontenc}

\DeclareRobustCommand{\VAN}[3]{#2}
\let\VANthebibliography\thebibliography
\def\thebibliography{\DeclareRobustCommand{\VAN}[3]{##3}\VANthebibliography}

\usepackage{graphicx}	
\usepackage{amsmath}	
\usepackage{multicol}
\usepackage{subcaption}

\title[Outflow of J1439-0106]{VLT/UVES Observation of the Outflow in Quasar SDSS J1439-0106}

\author[D. Byun et al.]{
Doyee Byun,$^{1}$\thanks{E-mail: dbyun@vt.edu}
Nahum Arav,$^{1}$
Andrew Walker$^{1}$
\\
$^{1}$Virginia Tech, Blacksburg, Virginia, USA
}

\date{Accepted XXX. Received YYY; in original form ZZZ}

\pubyear{2022}

\begin{document}
\label{firstpage}
\pagerange{\pageref{firstpage}--\pageref{lastpage}}
\maketitle

\begin{abstract}
We analyze the VLT/UVES spectrum of the quasar SDSS J143907.5-010616.7, retrieved from the UVES Spectral Quasar Absorption Database. We identify two outflow systems in the spectrum: a mini broad absorption line (mini-BAL) system and a narrow absorption line (NAL) system. We measure the ionic column densities of the mini-BAL ($v=-1550$ km s$^{-1}$) outflow, which has excited state absorption troughs of \ion{Fe}{ii}. We determine that the electron number density $\log{n_e}=3.4^{+0.1}_{-0.1}$ based on the ratios between the excited and ground state abundances of \ion{Fe}{ii}, and find the kinetic luminosity of the outflow to be $\lesssim 0.1 \%$ of the quasar's Eddington luminosity, making it insufficient to contribute to AGN feedback.
\end{abstract}




\section{Introduction}

Quasar outflows are often found in the spectra of quasars ($\lesssim 40\%$) as blueshifted absorption troughs relative to the rest frame of  the quasars \citep[][]{2003AJ....125.1784H,2008ApJ...672..108D,2008MNRAS.386.1426K}. Often invoked as potential contributors to AGN feedback, analysis of these outflows can provide us with insight into galaxy evolution \citep[e.g.][]{1998A&A...331L...1S,2018ApJ...857..121Y,2021ApJ...919..122V,2022SciA....8.3291H}. The outflows must have a kinetic luminosity ($\dot{E}_k$) of at least $\sim0.5\%$ \citep{2010MNRAS.401....7H} or perhaps as much as $\sim5\%$ \citep{2004ApJ...608...62S} of the quasar's Eddington luminosity ($L_\text{Edd}$), depending on the theoretical model, to be contributors of AGN feedback. Several outflows with sufficient $\dot{E}_k$ have been found in past studies \citep[e.g.][]{2009ApJ...706..525M,2013MNRAS.436.3286A,2015MNRAS.450.1085C,2018ApJ...866....7L,2019ApJ...876..105X,2020ApJS..247...39M,2022Byun,2022arXiv220311964C}.

Crucial to the process of finding a quasar outflow's kinetic luminosity is finding its mass flow rate ($\dot{M}$), which is dependent on its hydrogen column density ($N_H$), ionization parameter ($U_H$), and electron number density ($n_e$) \citep{2012ApJ...758...69B}. Analysis using this method has been conducted in the past (e.g. \citealt{2001ApJ...548..609D,2001ApJ...550..142H,2018ApJ...858...39X,2020ApJS..247...37A}; Walker et al. 2022, submitted). The value of $n_e$ can be found by calculating the ratios between the excited and resonance state column densities of ions \citep{2018ApJ...857...60A}. This paper presents the determination of $\dot{E}_k$ of one of the outflow components of the quasar SDSS J143907.50-010616.7 (hereafter J1439-0106), based on the normalized VLT/UVES spectrum acquired from the Spectral Quasar Absorption Database (SQUAD) published by \citet{Murphy2019}. Similar analysis using data from this database has been conducted in previous studies (\citealt{2022Byun}; Byun et al. 2022, submitted; Walker et al. 2022, submitted).

The UVES data of J1439-0106 was collected as part of the programs 081.B-0285(A) and 083.B-0604(A), and has been added to the SQUAD database complied by \citet{Murphy2019}. From the normalized spectrum, we identify two distinct absorption outflow systems, which we label here as the mini broad absorption line (mini-BAL) system S1, and the narrow absorption line (NAL) system S2, of which we find S1 suitable for our analysis thanks to the presence of excited state absorption troughs.

This paper is structured as follows. Section \ref{sec:observation} describes the observation of J1439-0106, as well as the method we used to retrieve its spectral data. Section \ref{sec:analysis} discusses the measurement of ionic column densities of S1, as well as the determination of $N_H$, $U_H$, and $n_e$. Section \ref{sec:results} shows the resulting calculation of $\dot{M}$ and $\dot{E}_k$, as well as its ratio with $L_\text{Edd}$. Section \ref{sec:discussion} provides a discussion of these results, as well as a comparison with previous work. Section \ref{sec:conclusion} summarizes and concludes the paper. We adopt a cosmology of $h=0.696$, $\Omega_m=0.286$, and $\Omega_\Lambda=0.714$ \citep{Bennett_2014}. We use the Python astronomy package Astropy \citep{astropy:2013,astropy:2018} for cosmological calculations.

\section{Observation and Data Acquisition}
\label{sec:observation}
\begin{figure*}
	\includegraphics[width=\linewidth]{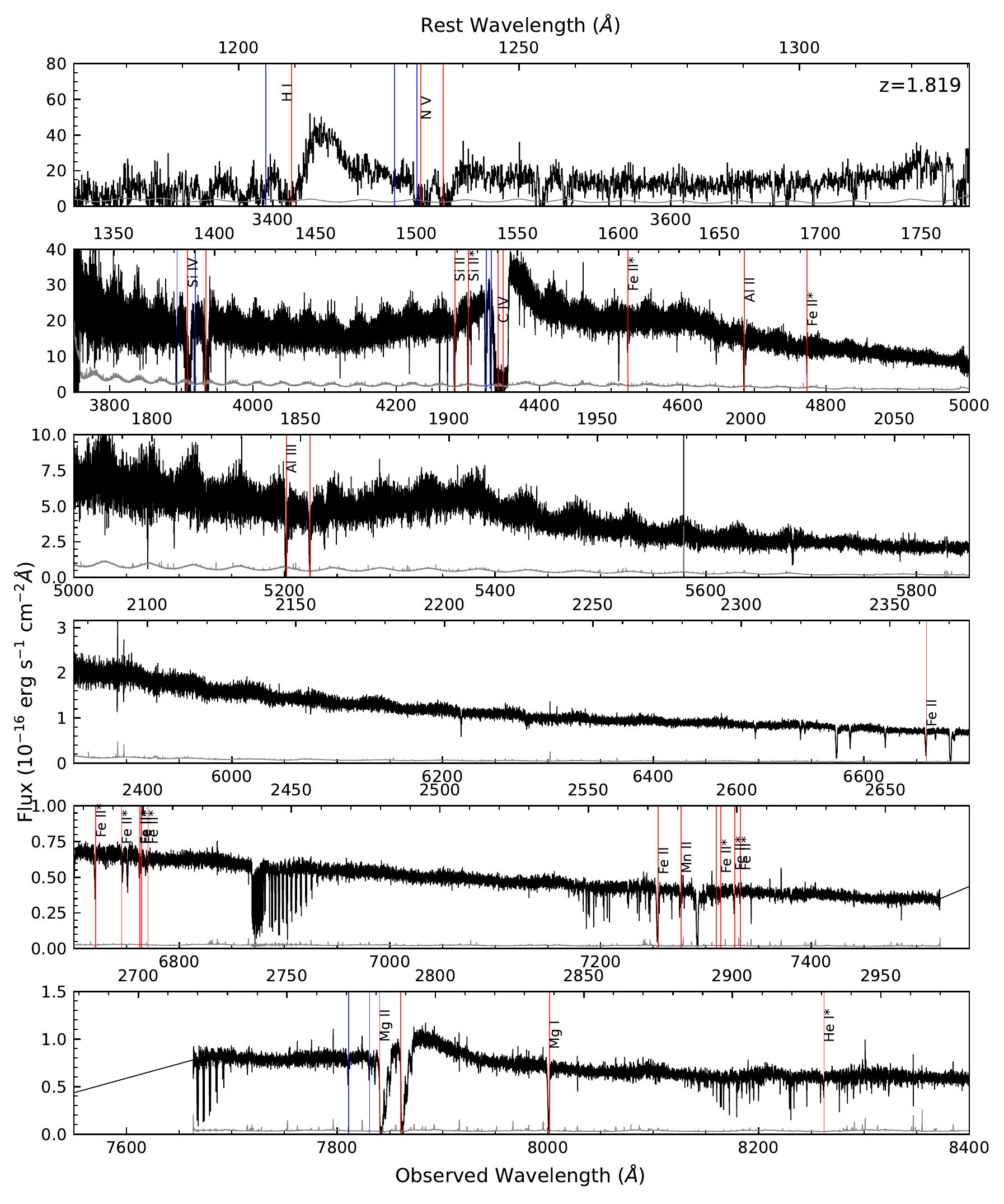}
    \caption{The UVES spectrum of J1439-0106. The normalized spectrum was multiplied by the continuum model by \citet{Murphy2019}, and scaled to match the flux of the SDSS spectrum at observed wavelength 6500\AA. Red vertical lines mark the absorption features of S1, and the blue vertical lines mark the features of S2. While in S2, troughs of \ion{C}{iv}, \ion{Si}{iv}, \ion{Mg}{ii}, \ion{H}{i}, and \ion{N}{v} are detected, they are shallower and narrower than those of S1. The first panel has been binned by 10 pixels for clarity.}
    \label{fig:fluxplot}
\end{figure*}

J1439-0106 (J2000: RA=14:39:07.5, DEC=01:06:16.7, z=1.819) was observed with VLT/UVES on 1 May, 2008 as part of the program 081.B-0285(A), and on 16 April, 2009 as part of 083.B-0604(A), with minimal variability between the two epochs. The combined spectral data, covering wavelengths between 3284--9466\AA, was normalized by the quasar's continuum and emission, and added to the SQUAD database by \citet{Murphy2019}. The spectrum is shown in Fig.~\ref{fig:fluxplot}. The object was also observed in 15 May, 2002 as part of the Sloan Digital Sky Survey (SDSS), the spectrum of which we use to calibrate the flux shown in Fig.~\ref{fig:fluxplot}, as well as find the bolometric and Eddington luminosities of the quasar.

From the UVES spectrum, we identify two different absorption outflow systems, mini-BAL S1 ($v\approx-1550$ km s$^{-1}$) and NAL S2 ($v\approx-2750$ km s$^{-1}$). We focus on S1 in the analysis of this paper, as it shows troughs of several \ion{Fe}{ii} excited lines, allowing us to find the outflow distance from the source, and by extension, the mass flow rate.

\section{Analysis}
\label{sec:analysis}
\subsection{Ionic Column Densities}
\begin{figure*}
    \centering
    \begin{multicols}{3}
    \subcaptionbox{\ion{H}{i}\label{fig:HI}}{\includegraphics[width=\linewidth]{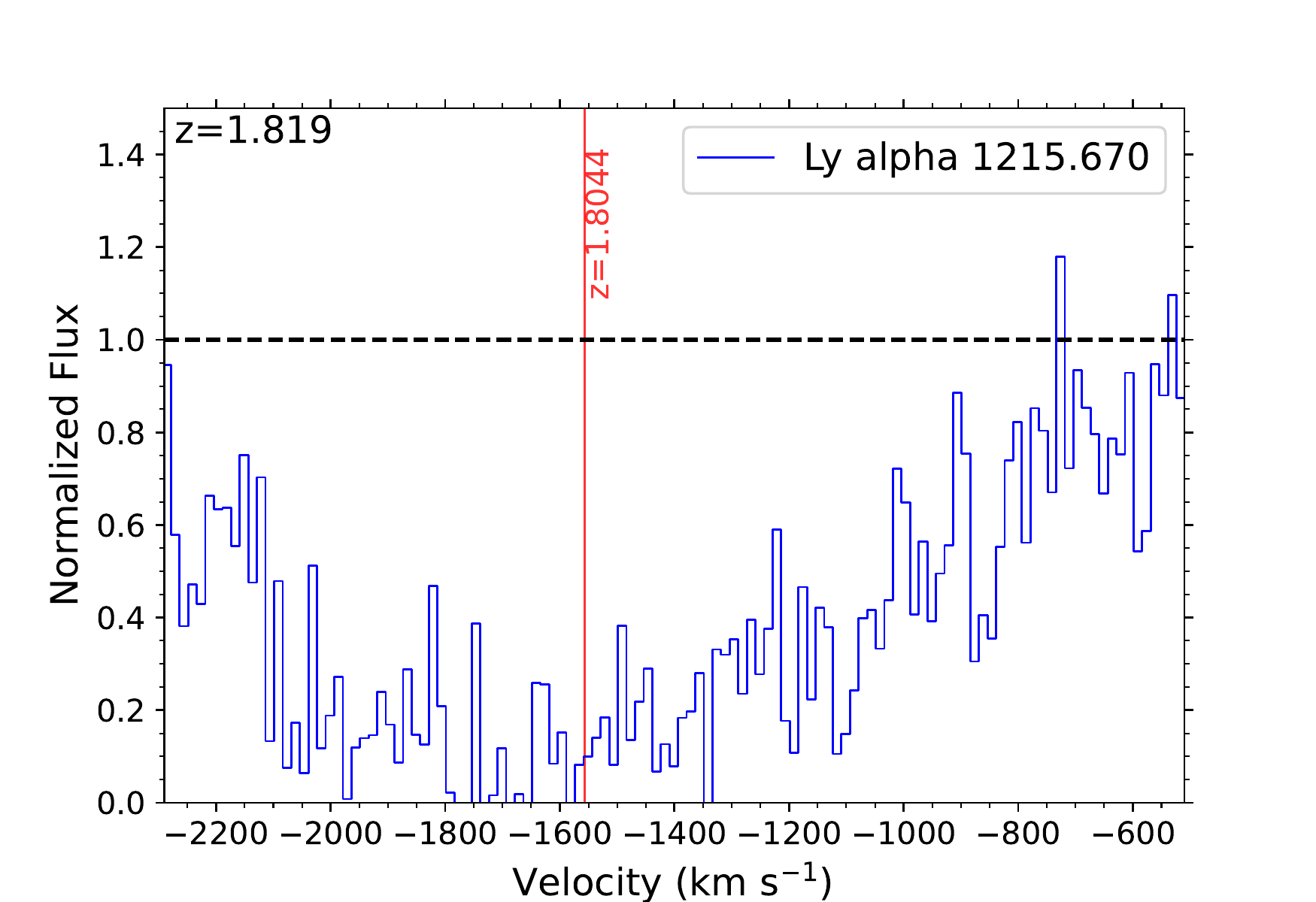}}\par
    \subcaptionbox{\ion{He}{i}*\label{fig:NV}}{\includegraphics[width=\linewidth]{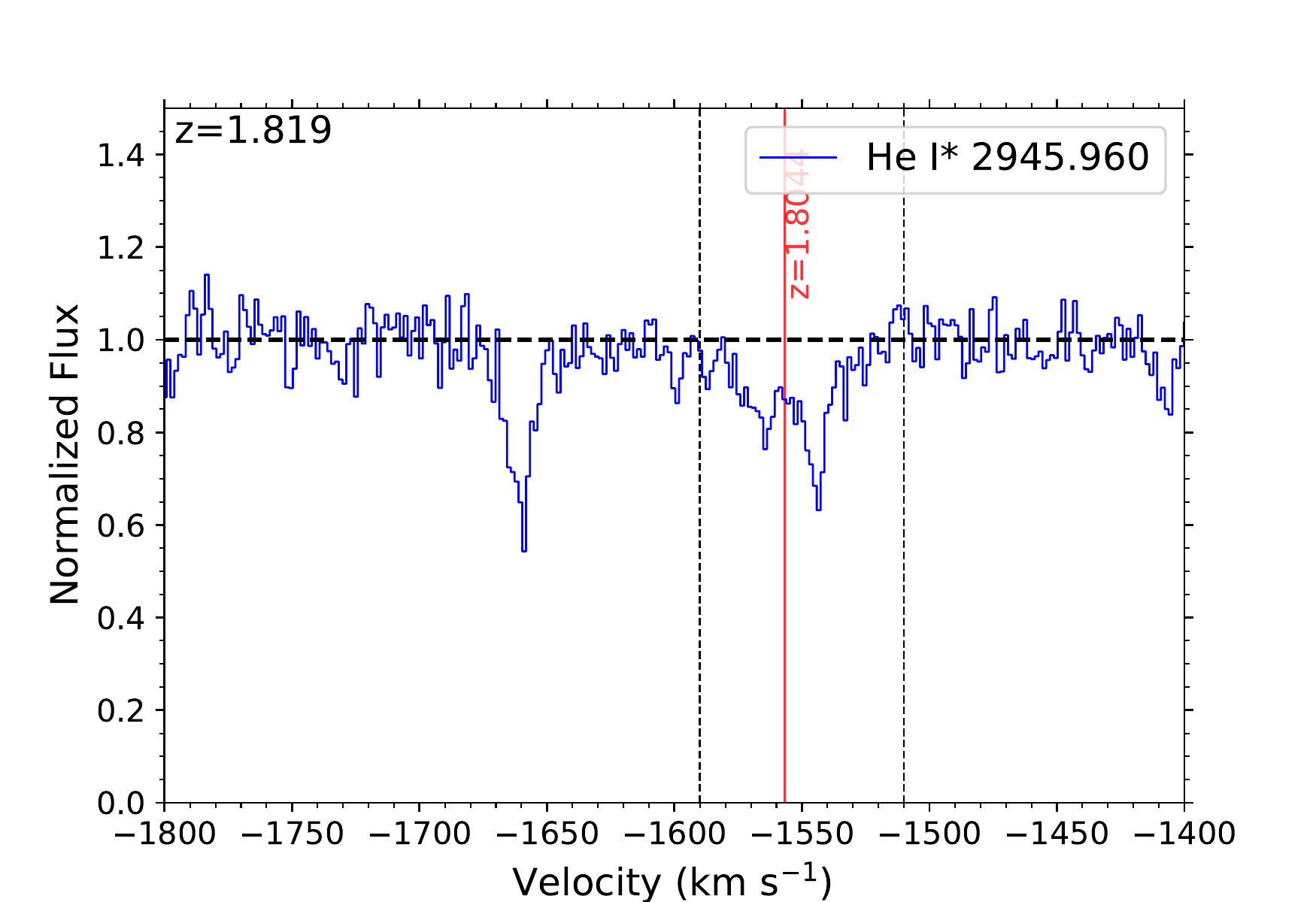}}\par
    \subcaptionbox{\ion{C}{iv}\label{fig:PV}}{\includegraphics[width=\linewidth]{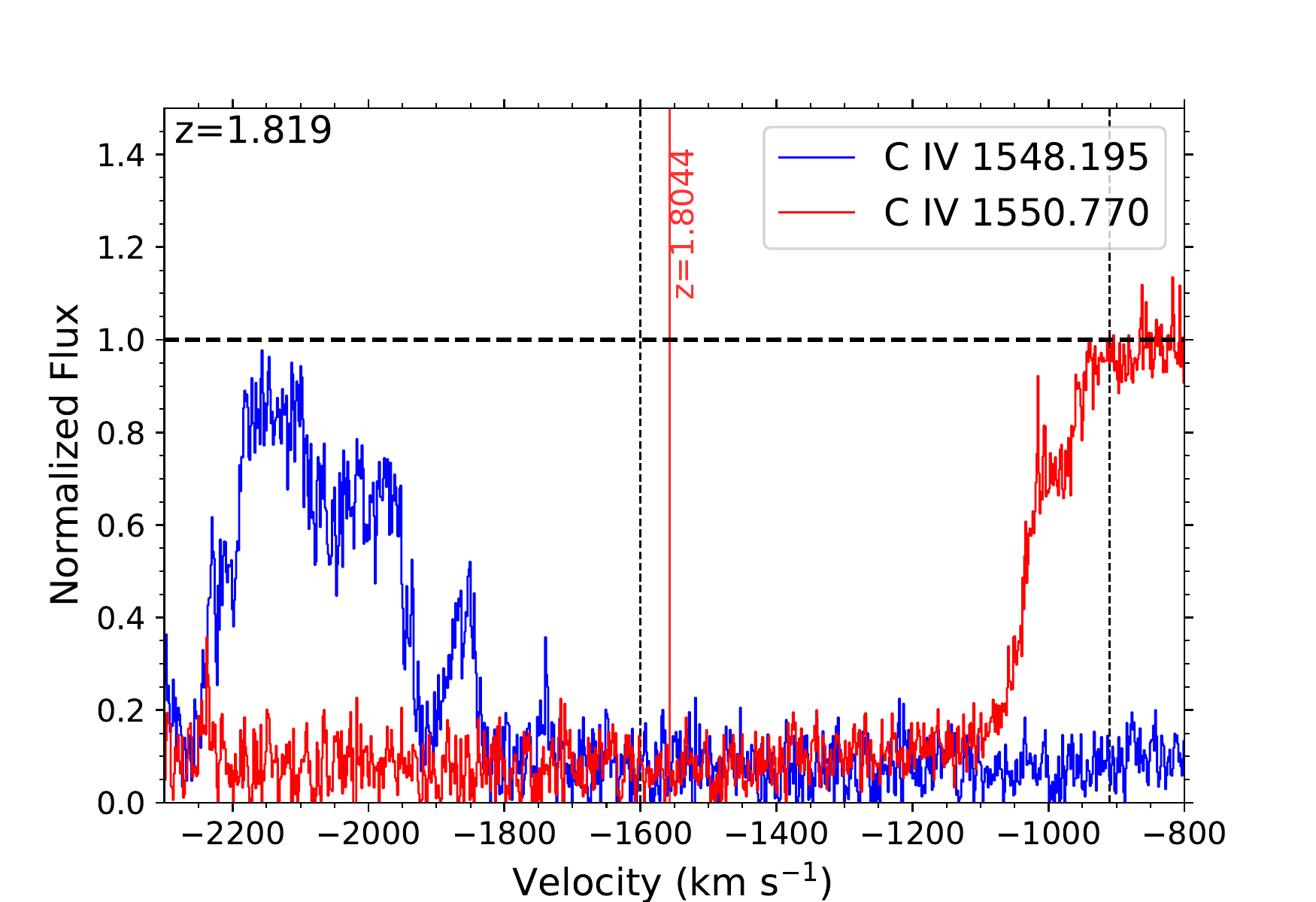}}\par
    \end{multicols}
    \begin{multicols}{3}
    \subcaptionbox{\ion{N}{v}\label{fig:SVI}}{\includegraphics[width=\linewidth]{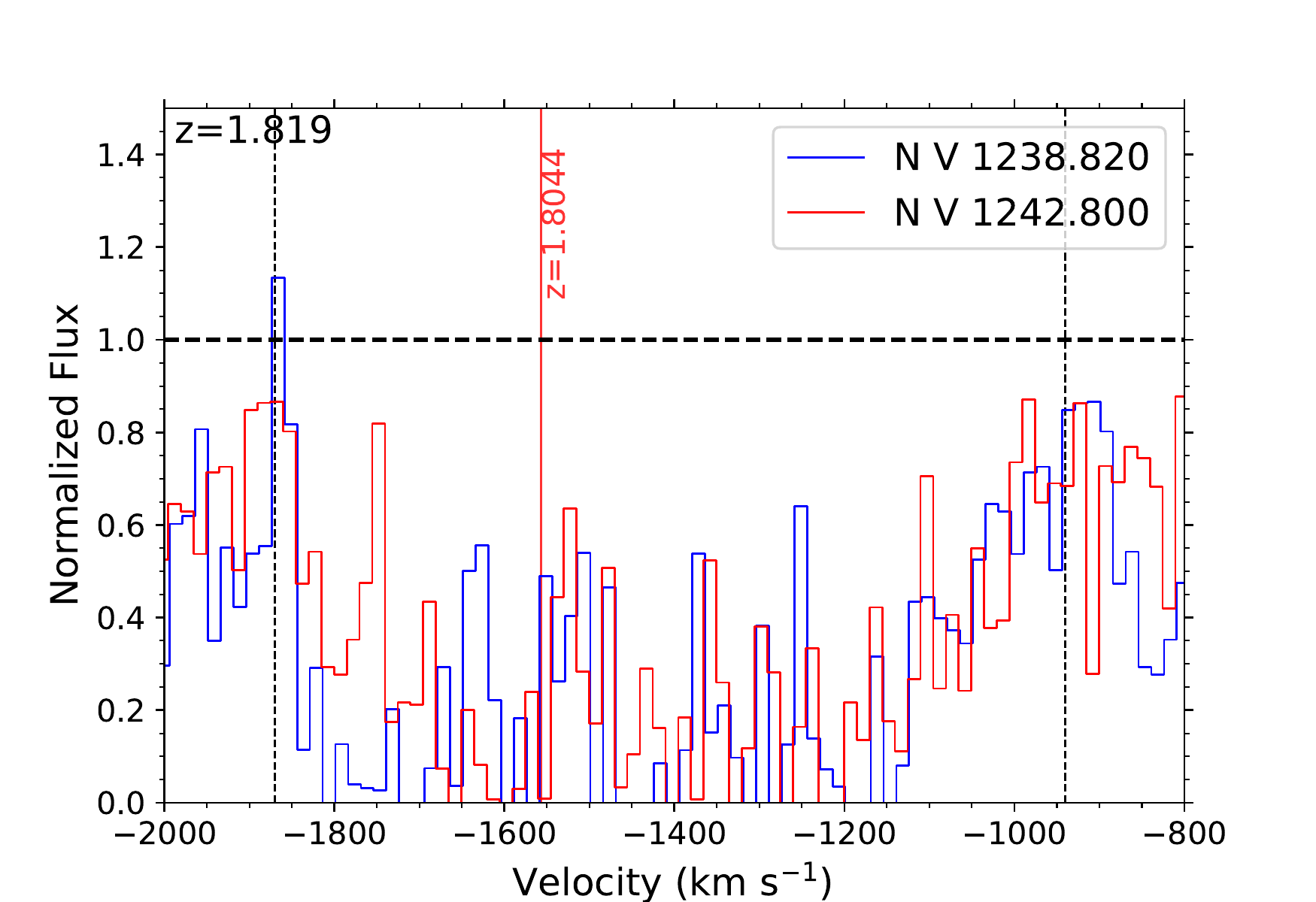}}\par
    \subcaptionbox{\ion{Mg}{i}\label{fig:CIII}}{\includegraphics[width=\linewidth]{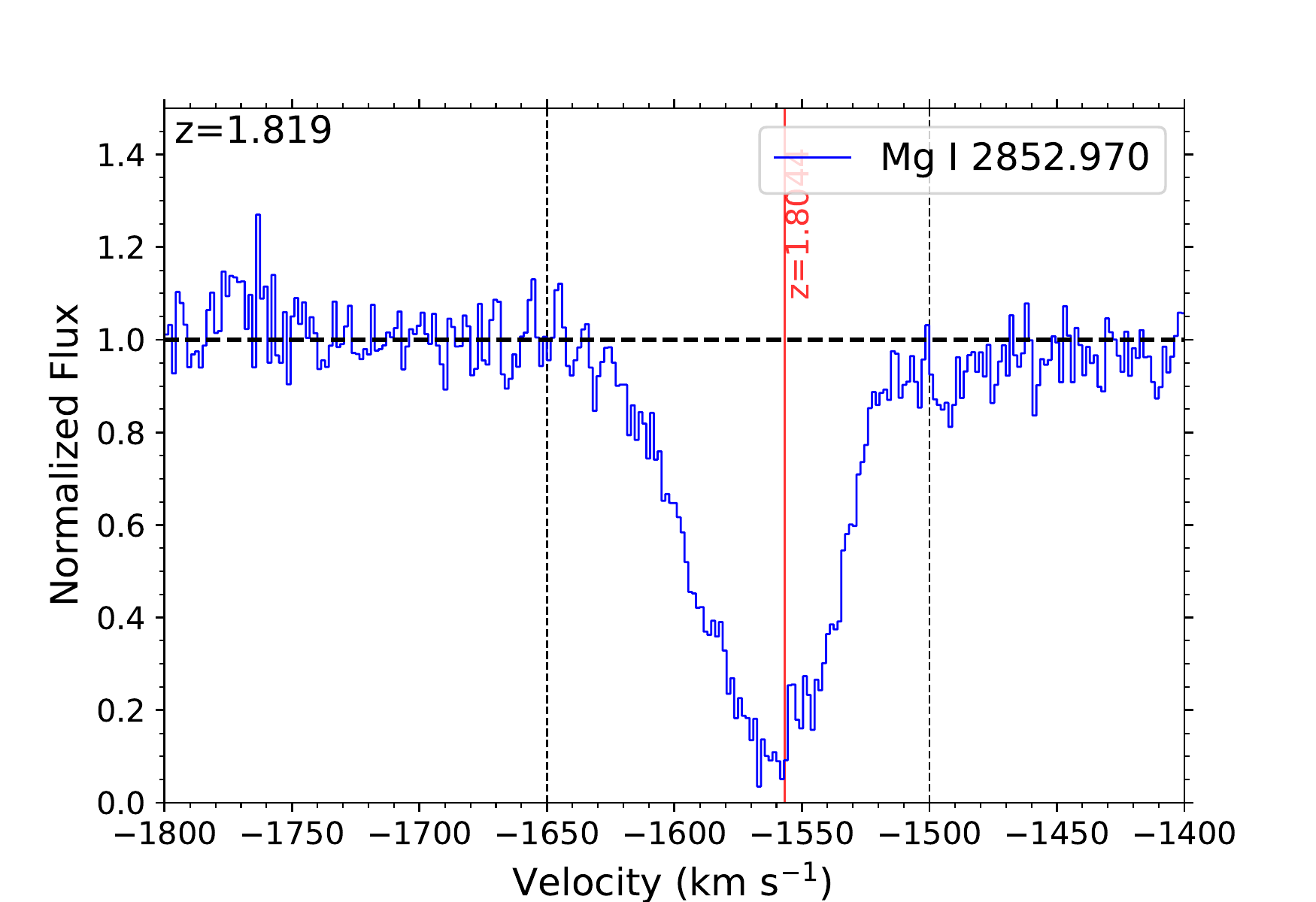}}\par
    \subcaptionbox{\ion{Mg}{ii}\label{fig:CIV}}{\includegraphics[width=\linewidth]{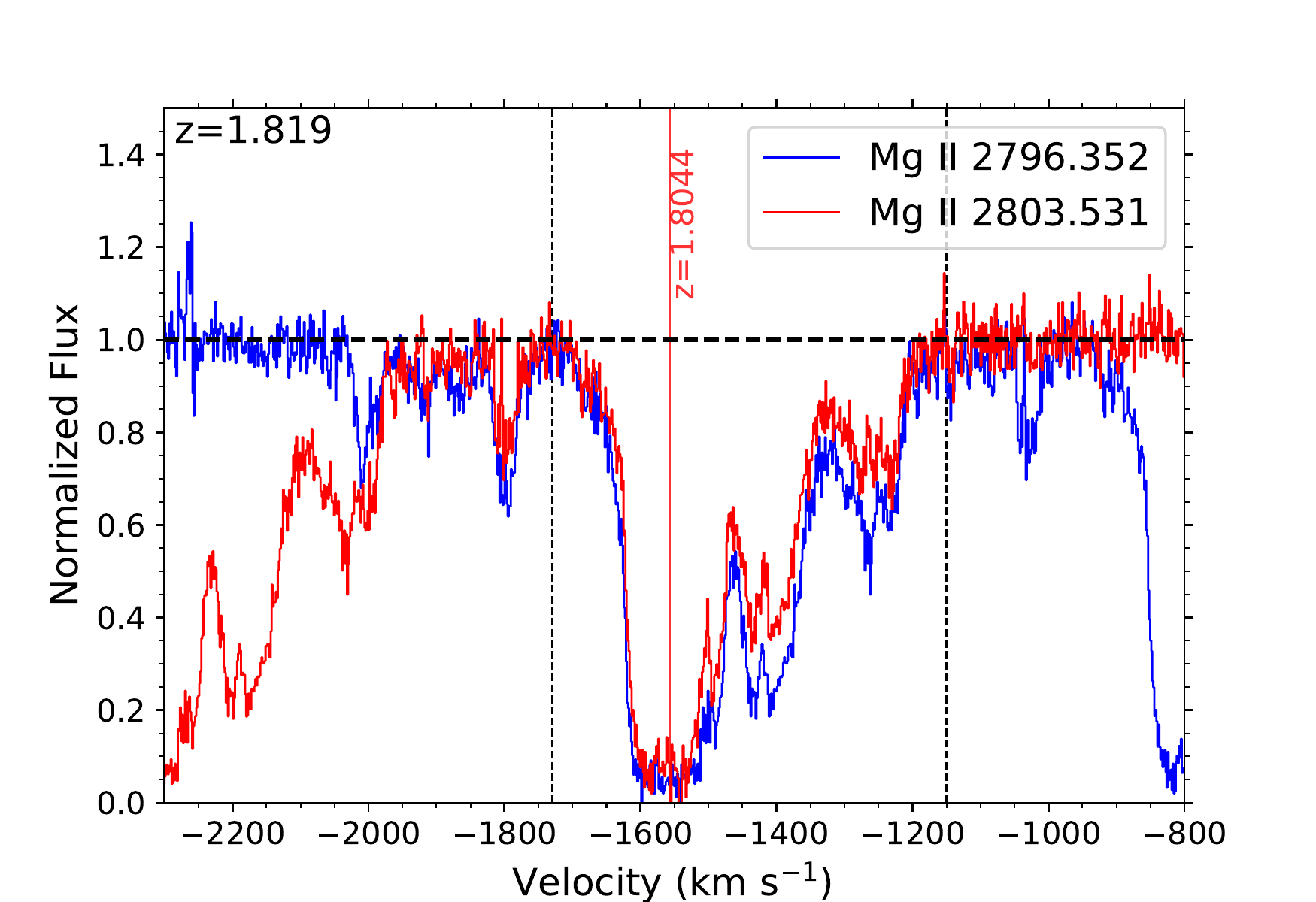}}\par
    \end{multicols}
    \begin{multicols}{3}
    \subcaptionbox{\ion{Al}{ii}\label{fig:SiIV}}{\includegraphics[width=\linewidth]{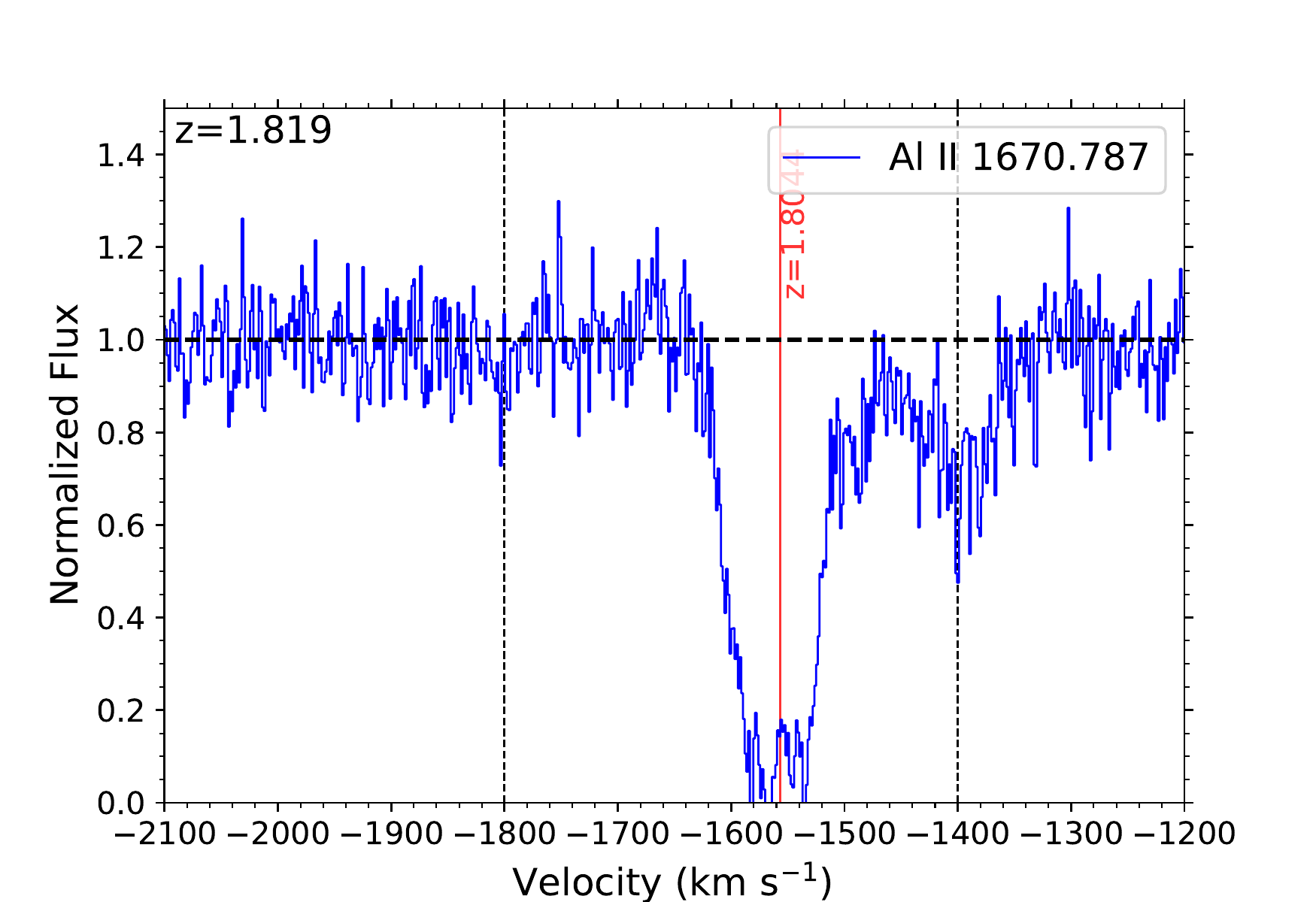}}\par
    \subcaptionbox{\ion{Al}{iii}\label{fig:AlII}}{\includegraphics[width=\linewidth]{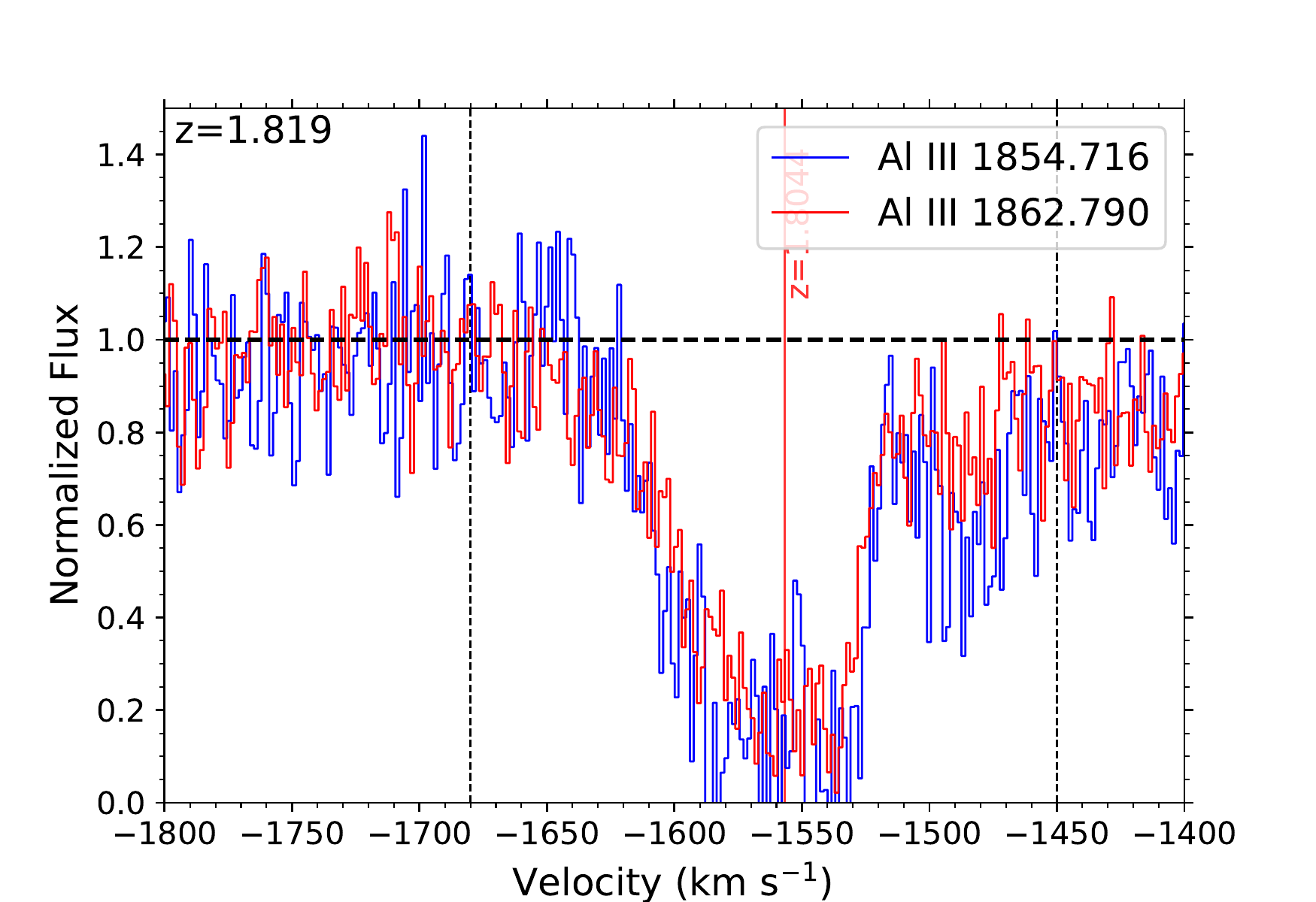}}\par
    \subcaptionbox{\ion{Si}{ii}\label{fig:AlIII}}{\includegraphics[width=\linewidth]{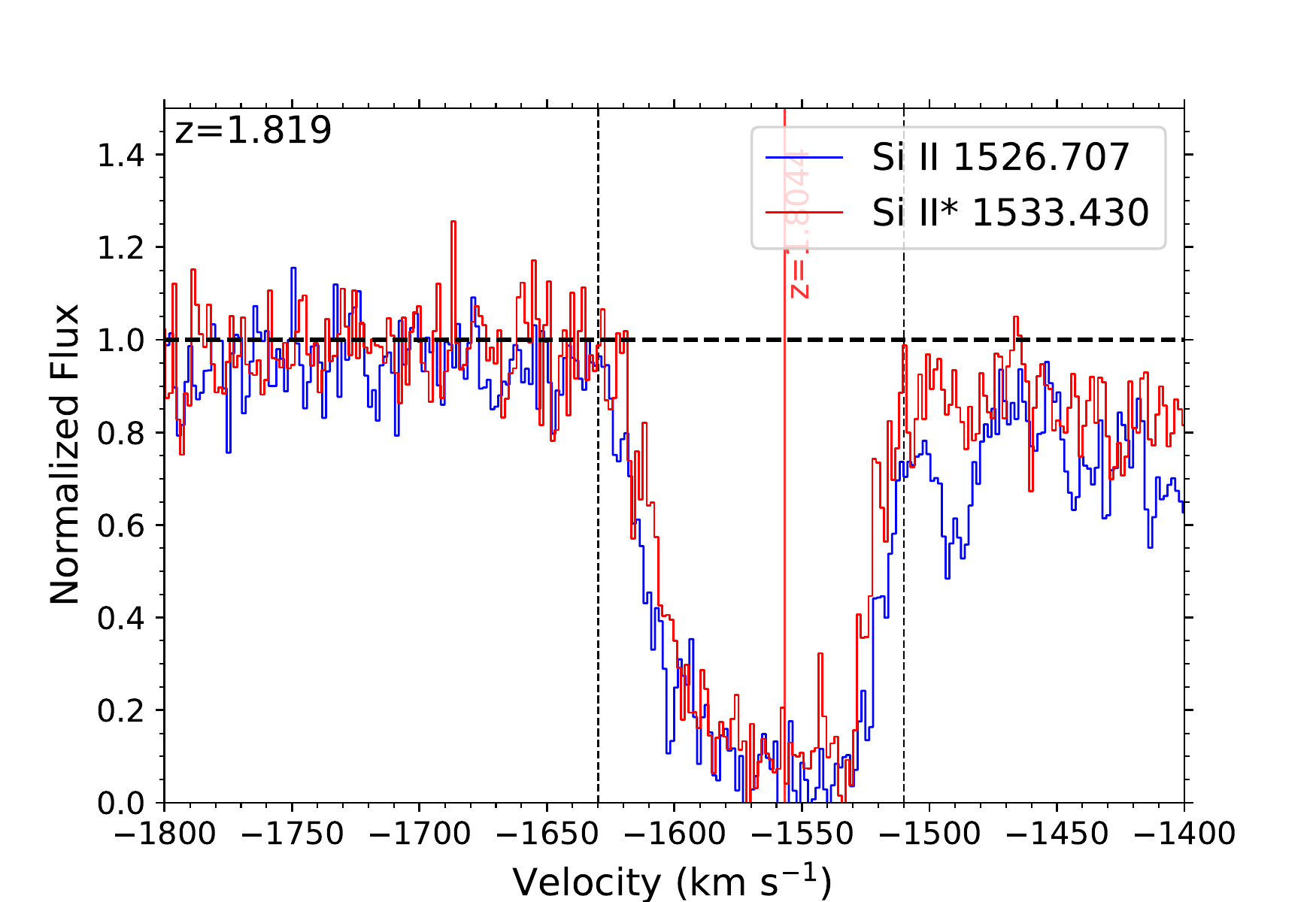}}\par
    \end{multicols}
    \begin{multicols}{3}
    \subcaptionbox{\ion{Si}{iv}\label{fig:SiIV}}{\includegraphics[width=\linewidth]{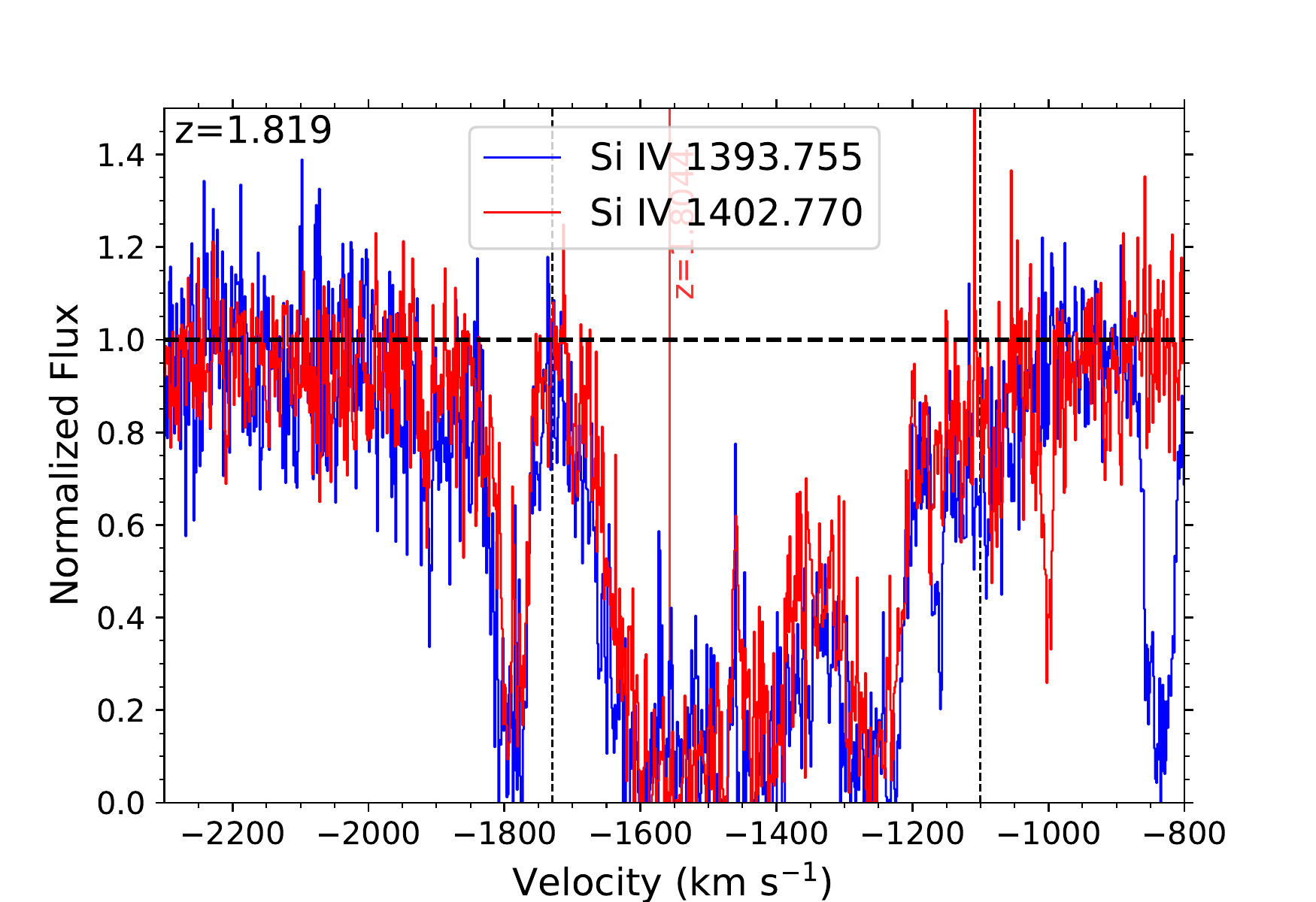}}\par
    \subcaptionbox{\ion{Mn}{ii}\label{fig:AlII}}{\includegraphics[width=\linewidth]{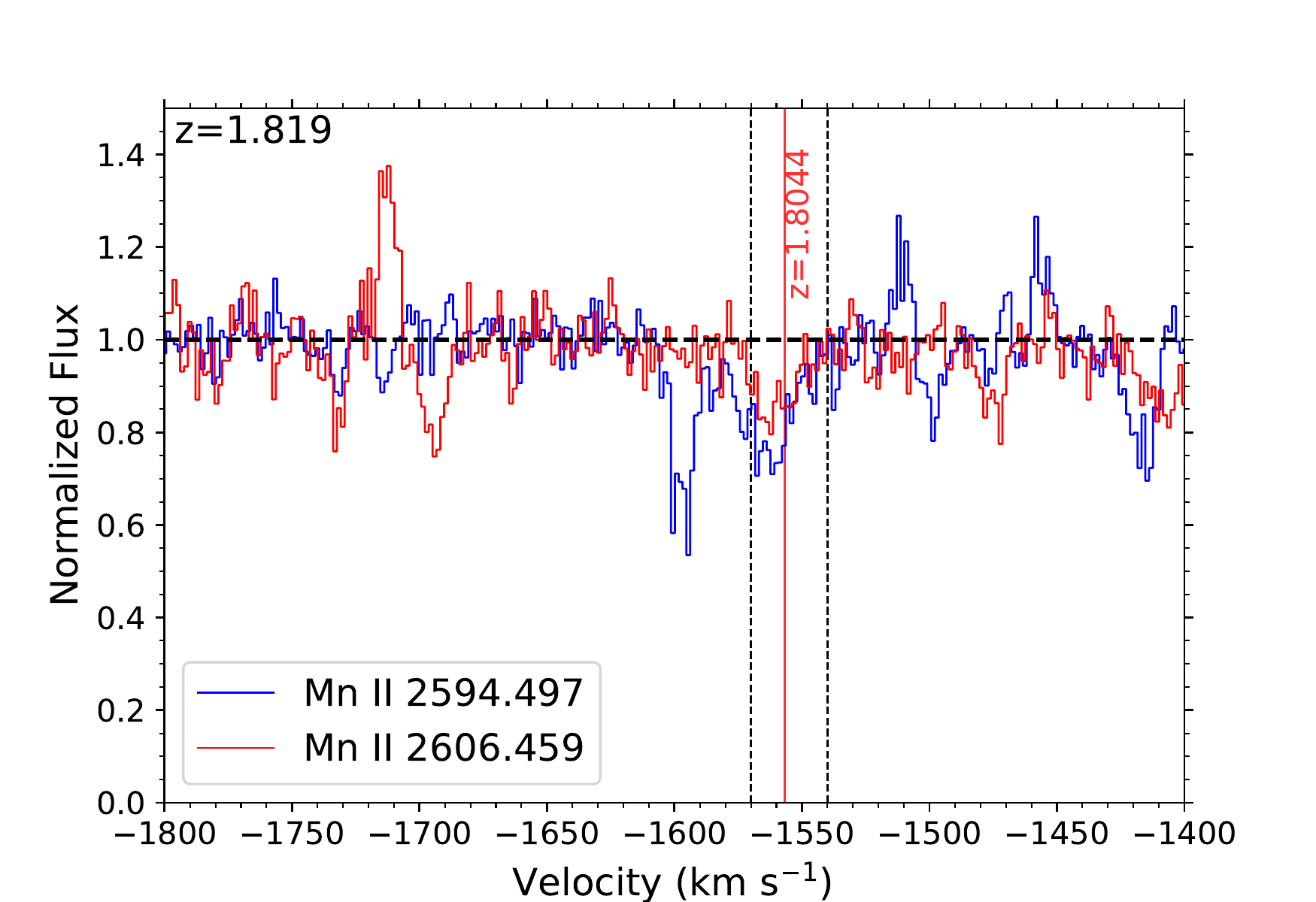}}\par
    \subcaptionbox{\ion{Fe}{ii}\label{fig:AlIII}}{\includegraphics[width=\linewidth]{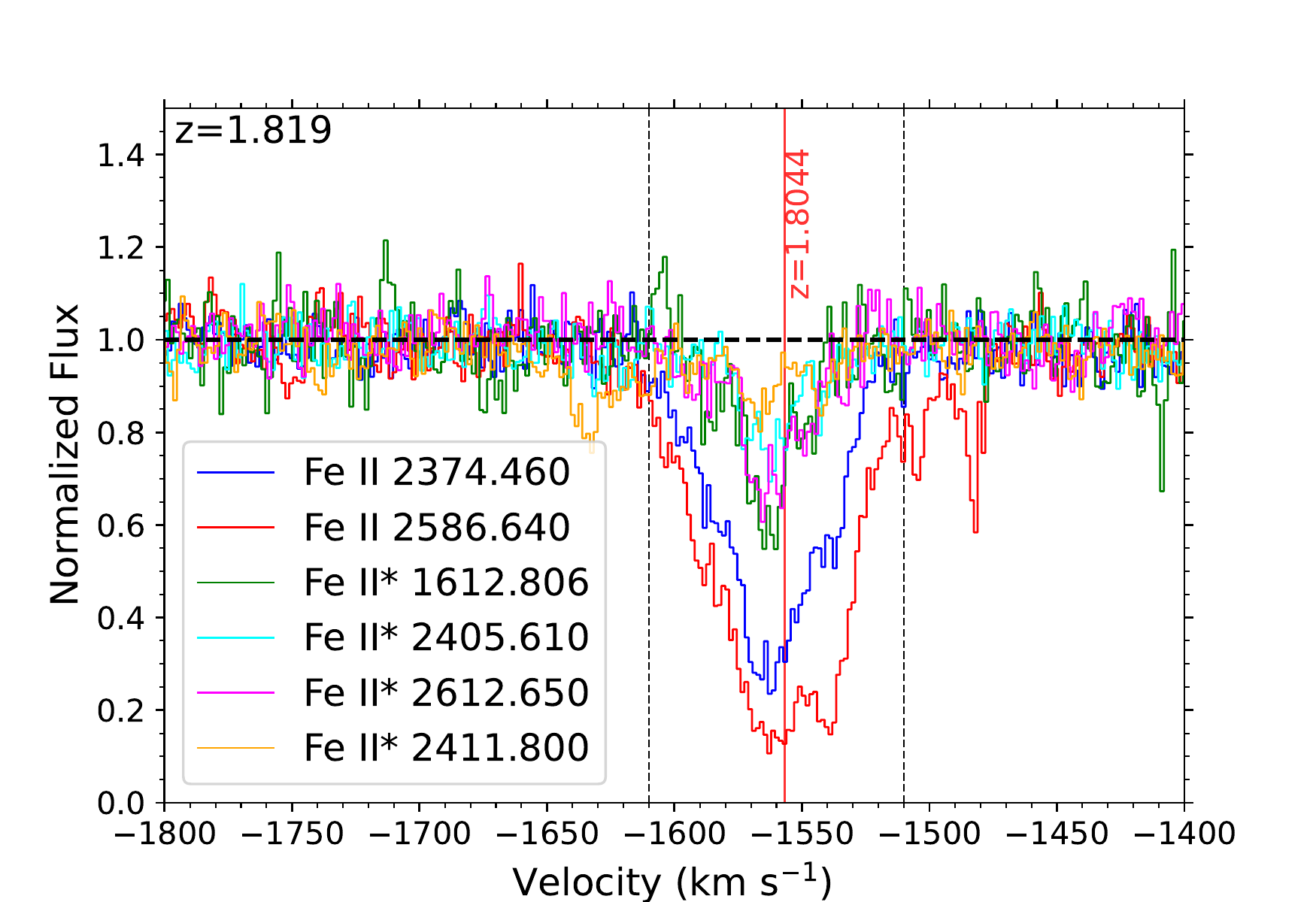}}\par
    \end{multicols}
\caption{Absorption troughs of the outflow system  S1 plotted in velocity space. The horizontal dashed line indicates the continuum level, while the vertical red line shows the velocity of S1. The vertical dotted lines show the integration range used to find the column densities.}
\label{fig:vcut}
\end{figure*}

Finding the ionic column densities ($N_{ion}$) of S1 is crucial to determining the energetics parameters of the outflow system. In order to measure the column densities, we use the systemic redshift of the quasar to convert the spectrum from wavelength space to velocity space, as shown in Fig.~\ref{fig:vcut}. We then use two different methods to find the ionic column densities, assuming either an apparent optical depth (AOD) of a uniform outflow \citep{Savage1991}, or partial covering (PC) based on a velocity dependent covering factor \citep{1997ASPC..128...13B,1999ApJ...524..566A,Arav1999}.

The AOD method and PC method have different advantages over one another, with the PC method being particularly more helpful in finding more accurate column densities for ions with absorption doublets or multiplets, while the AOD method yields lower limits to the column densities \citep{DeKool2002,2005ApJ...620..665A,2011ApJ...739....7E,2012ApJ...751..107B}. The differences between the two methods is explained in further detail in Section 3.1 of \citet{2022Byun}.

We choose our integration range for each ion based on the visibility of absorption troughs, as shown in Fig.~\ref{fig:vcut}. In cases were the red and blue troughs of a doublet are blended (e.g. \ion{C}{iv}), we choose a range in which the red and blue troughs are not overlapping, in order to gain a lower limit of the column density. The measured column densities are shown in Table~\ref{table:coldensity}. Note that we add a 20\% error in quadrature to account for the uncertainty in the continuum model \citep{2018ApJ...858...39X}.

\begin{table}
	\centering
	\caption{J1439-0106 outflow column densities from UVES observations. The values are in units of $10^{12}$ cm$^{-2}$.}
	\label{table:coldensity}
	\renewcommand{\arraystretch}{1.4}
	\begin{tabular}{lcccc} 
	\hline\hline
	Troughs&AOD&PC&Adopted\\
	\hline
	\multicolumn{4}{c}{S1, $v=-1550\text{ km s}^{-1}$}\\
	\hline
    \text{\ion{H}{i}} &$1450_{-40}^{+80}$  &&$>1450_{-290}$\\
    \text{\ion{He}{i}*} & $100_{-6}^{+6}$&&$>100_{-20}$\\
    \text{\ion{C}{iv}} &$3440_{-40}^{+50}$  &&$>3440_{-690}$\\
    \text{\ion{N}{v}} &$3370_{-140}^{+220}$&&$>3370_{-690}$\\
    \text{\ion{Mg}{i}} & $7.8_{-0.2}^{+0.2}$&&$>8_{-1.6}$\\
    \text{\ion{Mg}{ii}} & $196_{-3}^{+4}$&$250_{-4}^{+70}$&$250_{-50}^{+80}$\\
    \text{\ion{Al}{ii}} & $25_{-1}^{+2}$&&$>25_{-5}$\\
    \text{\ion{Al}{iii}} & $96_{-3}^{+3}$&&$>100_{-20}$\\
    \text{\ion{Si}{ii} total} & $720_{-20}^{+30}$&&$>720_{-100}$\\
    \text{\ion{Si}{ii} 0} & $410_{-15}^{+25}$&&\\
    \text{\ion{Si}{ii}*} & $310_{-10}^{+15}$&&\\
    \text{\ion{Si}{iv}} & $850_{-20}^{+40}$&&$>850_{-170}$\\
    \text{\ion{Mn}{ii}} & $2.7_{-0.3}^{+0.3}$&$2.8_{-0.2}^{+0.2}$&$2.8_{-0.6}^{+0.6}$\\
    \text{\ion{Fe}{ii} total} &&&$>430_{-60}$\\
    \text{\ion{Fe}{ii} 0} & $230_{-4}^{+4}$&$270_{-5}^{+5}$&$270_{-50}^{+50}$\\
    \text{\ion{Fe}{ii}* 385} & $16_{-3}^{+3}$&$21_{-1}^{+3}$&$21_{-4}^{+5}$\\
    \text{\ion{Fe}{ii}* 668} & $14_{-4}^{+4}$&$18_{-4}^{+6}$&$18_{-6}^{+7}$\\
    \text{\ion{Fe}{ii}* 862} & $7_{-1}^{+1}$&&$7_{-1.6}^{+1.5}$\\
    \text{\ion{Fe}{ii}* 977} & $3.8_{-0.3}^{+0.4}$&&$3.8_{-0.8}^{+0.8}$\\
    \text{\ion{Fe}{ii}* 1873} & $100_{-10}^{+10}$&$110_{-10}^{+10}$&$110_{-20}^{+20}$\\
		\hline
	\end{tabular}
\end{table}
\subsection{Photoionization Analysis}
\begin{figure*}
    \centering
    \begin{multicols}{2}
    \subcaptionbox{Solar metallicity\label{fig:solar}}{\includegraphics[width=\linewidth]{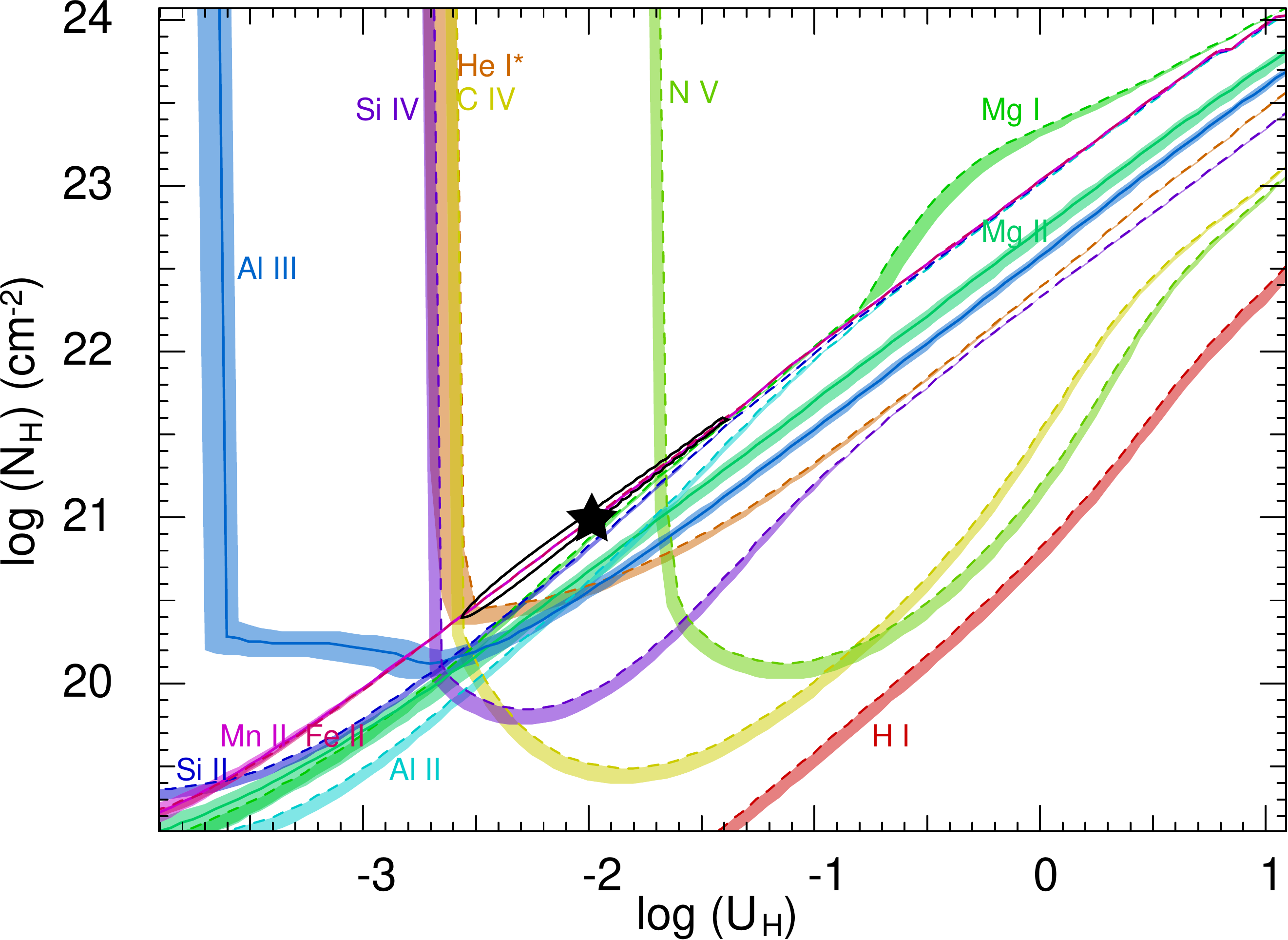}}\par
    \subcaptionbox{Super-solar metallicity (see text)\label{fig:super-solar}}{\includegraphics[width=\linewidth]{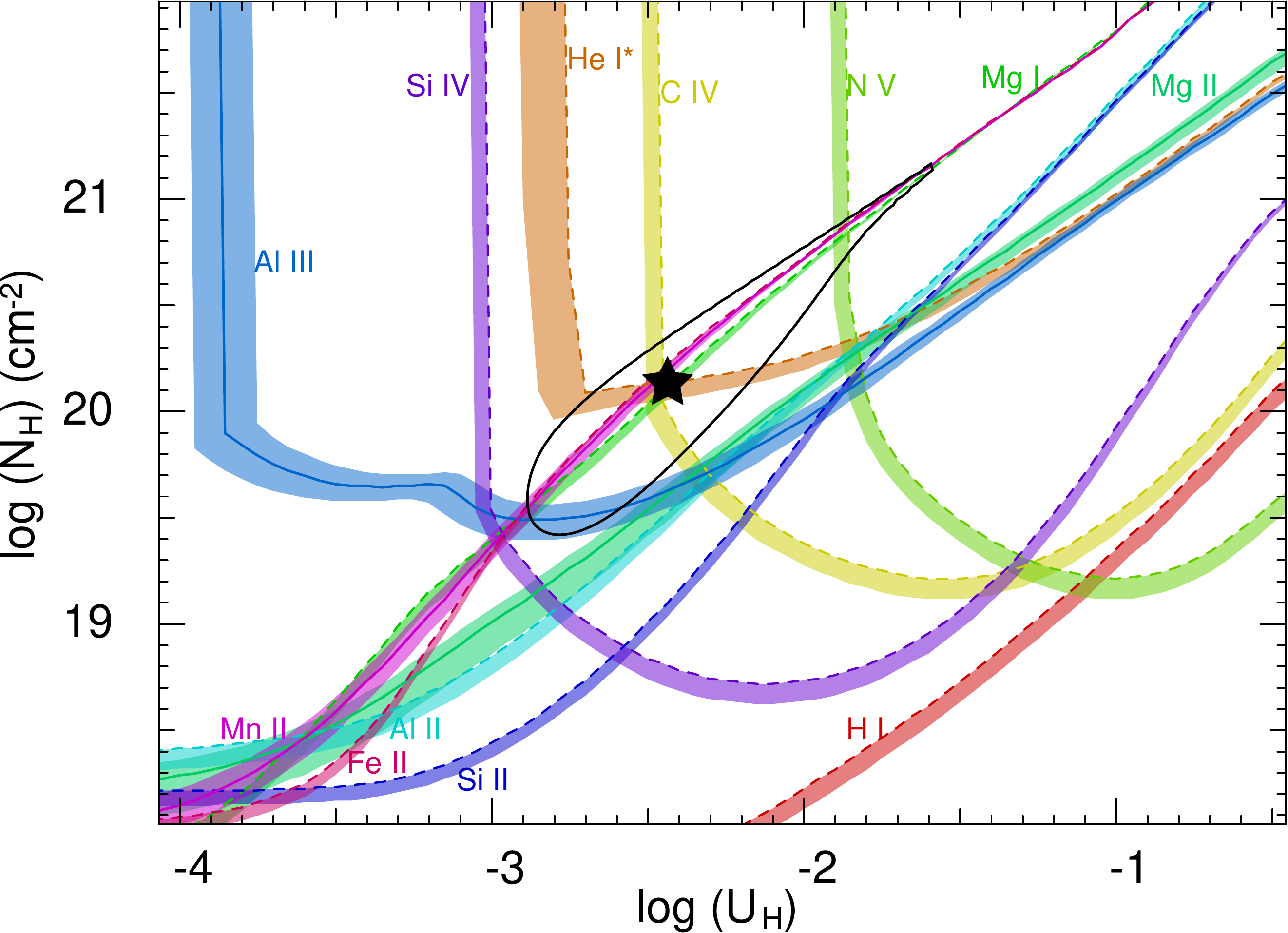}}\par
    \end{multicols}

\caption{Plots of $\log{N_H}$ vs. $\log{U_H}$ based on the ionic column densities of S1. The colored lines show the values of $N_H$ and $U_H$ allowed by the measured column density of each ion. Solid lines indicate measurements, while the dashed lines show lower limits from using the AOD method. The shaded bands attached to the lines represent the uncertainties in column density. The black stars indicate the solutions found via $\chi^2$ analysis, and the black ellipses represent the $1\sigma$ error.}
\label{fig:nvuplots}
\end{figure*}

With the ionic column densities found, we can use these measurements to find the hydrogen column density ($N_H$) and ionization parameter ($U_H$) of S1 \citep[e.g.][Walker et al. submitted]{2019ApJ...876..105X,2020ApJS..247...39M,2022Byun}. We use the spectral synthesis code Cloudy \citep[][version c17.00]{2017RMxAA..53..385F} to create a grid of simulated models based on varying values of $N_H$ and $U_H$, using the spectral energy distribution (SED) of quasar HE0238-1904 (hereafter HE0238) \citep{2013MNRAS.436.3286A}. Via $\chi^2$ analysis, we find the model with ionic column densities that best match the measured values, as shown in Fig.~\ref{fig:nvuplots}.  We use two different grids based on metallicity values, solar and super-solar \citep[$Z=4.68Z_\odot$][]{2008A&A...478..335B,2020ApJS..247...41M}, to find two different solutions, as previous studies show that the metallicities of outflows are between $Z_\odot$ and $5 Z_\odot$ \citep[e.g.][]{2006ApJ...646..742G,2007ApJ...658..829A,2020ApJS..247...41M}. The values of $N_H$ and $U_H$ are shown in Table~\ref{table:energetics}.
\subsection{Electron Number Density}
\begin{figure}
	\includegraphics[width=\linewidth]{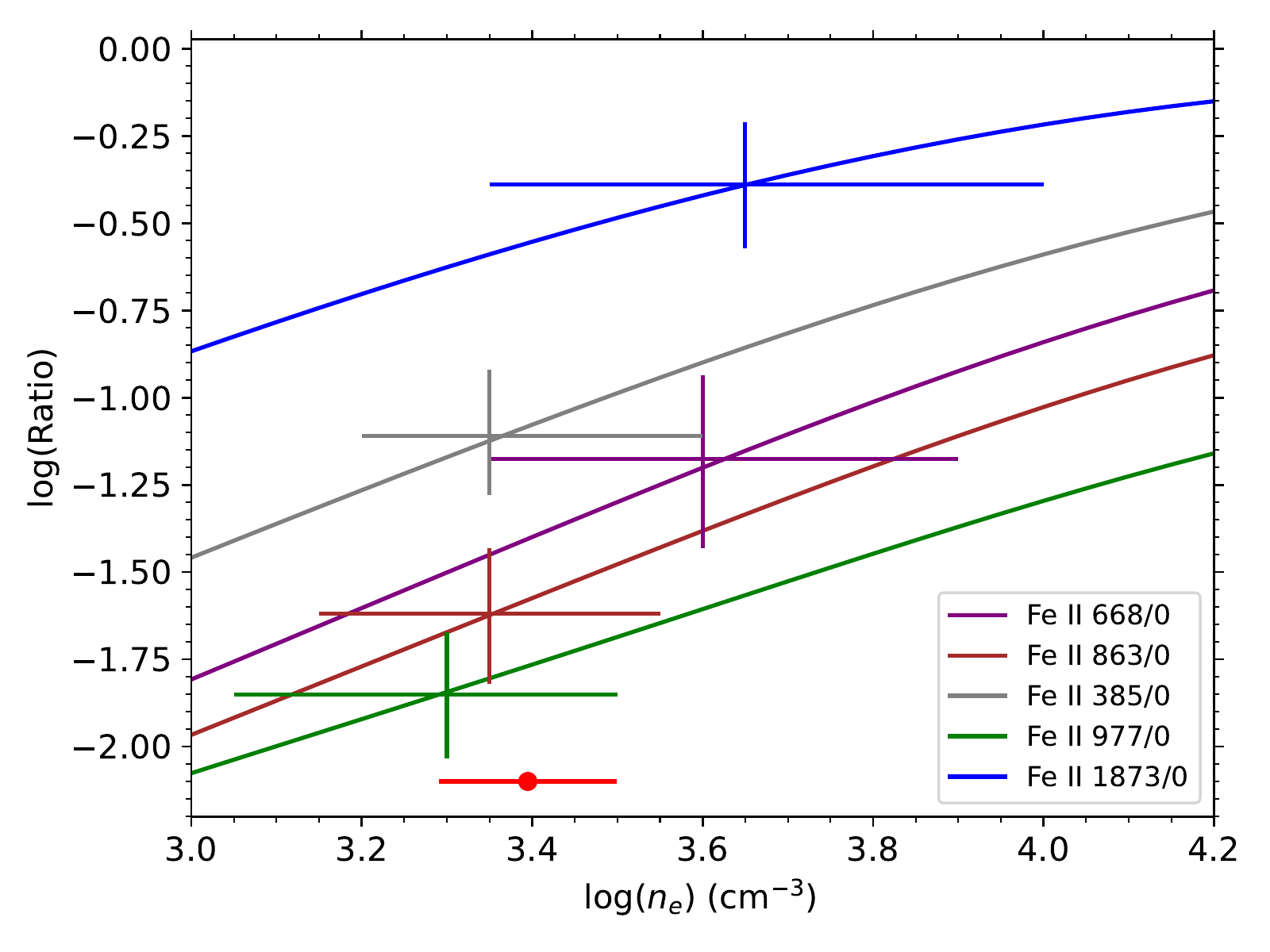}
    \caption{Log-log plot of the ratios between excited state \ion{Fe}{ii} and resonance state \ion{Fe}{ii} vs. electron number density. The curves show the theoretical relationship between the ratios and $n_e$ values, while the crosses show the measured ratios and their associated $n_e$. The red dot with error bars indicates the weighted mean of $\log{n_e}$. A temperature of 10,000 K is assumed.}
    \label{fig:ratioplot}
\end{figure}

Finding the distance of the outflow from its source is crucial to finding the mass flow rate, and by extension, the kinetic luminosity. This is done by finding the electron number density ($n_e$), which is measured by taking the ratios between excited and ground state column densities of ions \citep[e.g.][]{2009ApJ...706..525M,2022Byun}. The CHIANTI 9.0.1 Database \citep{1997A&AS..125..149D,Dere_2019} models the $n_e$ dependent ratios between different energy states based on collisional excitation, and can be used to find the value of $n_e$ from measured column densities. S1 shows absorption lines of five different excited states, as well as the ground state, of \ion{Fe}{ii}, and we find $n_e$ by finding the ratios between these excited states and the resonance state ($N(\ion{Fe}{ii}*)/N(\ion{Fe}{ii})$), as shown in Fig.~\ref{fig:ratioplot}. While we find troughs of \ion{Si}{ii} and \ion{Si}{ii}*, they are unreliable for finding $n_e$ due to the saturation of the troughs (see plot i in Fig.~\ref{fig:vcut}). We find the weighted mean of the $\log{n_e}$ values measured using the different excited states via the linear model method described by \citet{2003sppp.conf..250B}. This yields a value of $\log{n_e}=3.4^{+0.1}_{-0.1}$.
\section{Results}
\label{sec:results}
The distance of the outflow from the quasar can be found based on the definition of the ionization parameter:
\begin{equation}
    U_H\equiv\frac{Q_H}{4\pi R^2 n_H c}
\end{equation}
where $Q_H$ is the emission rate of hydrogen ionizing photons, $R$ is the distance from the source, $c$ is the speed of light,  and $n_H$ is the hydrogen number density. Once we find $Q_H$, we can use the values of $U_H$ and $n_e$ found in Section \ref{sec:analysis} to find $R$, as $n_e \approx 1.2 n_H$ in highly ionized plasma \citep{2006agna.book.....O}.

We find the value of $Q_H$ as follows, as per previous works \citep[e.g.][Byun et al. 2022, submitted; Walker et al. 2022, submitted]{2020ApJS..247...39M,2022Byun}. We scale the SED of HE0238 to match the continuum flux of J1439-0106 at observed wavelength $\lambda=6500$ \AA, from the SDSS observation of 15 May, 2002 ($F_\lambda=8.49^{+0.64}_{-0.64}\times10^{-17}$ erg s$^{-1}$ cm$^{-2}$ \AA$^{-1}$). We then integrated over the SED for energies over 1 Ryd, resulting in $Q_H=5.3^{+0.4}_{-0.4}\times10^{56}$ s$^{-1}$.

Once the distance is found, we can find the mass flow rate ($\dot{M}$) and kinetic luminosity ($\dot{E}_k$) as shown in the following \citep{2012ApJ...751..107B}:
\begin{equation}
    \dot{M} \simeq 4\pi\Omega R N_H \mu m_p v
\end{equation}
\begin{equation}
    \dot{E}_k\simeq\frac{1}{2}\dot{M}v^2
\end{equation}
where $\Omega$ is the fraction of the solid angle covered by the outflow, $\mu=1.4$ is the molecular weight, $m_p$ is the mass of a proton, and $v$ is the outflow velocity. We assume $\Omega=0.2$ based on ratio of quasars with \ion{C}{iv} BALs reported by \citet{2003AJ....125.1784H}. When propagating the uncertainties of the parameters, we take into account the positive correlation between $U_H$ and $N_H$ in the photoionization solutions (see Fig.~\ref{fig:nvuplots}) to avoid overestimating our errors (see Walker et al. (2022, submitted) for a detailed explanation). The values found for $\dot{M}$ and $\dot{E}_k$ are in Table~\ref{table:energetics}.
\begin{table}
	\centering
	\caption{Physical Properties of J1439-0106 Outflow.}
	\label{table:energetics}
	\begin{tabular}{lcc}
	\hline\hline
	Solution &\text{Solar}&\text{Super-solar}\\
	\hline
\vspace{-0.2cm}$log(N_{\text{H}})$&\\\vspace{-0.2cm}
&$20.99^{+0.61}_{-0.59}$&$20.13^{+1.03}_{-0.71}$\\
$[\text{cm}^{-2}]$&\\
\hline
\vspace{-0.2cm}$log(U_{\text{H}})$&\\\vspace{-0.2cm} &$-1.99^{+0.60}_{-0.58}$&$-2.44^{+0.85}_{-0.45}$\\
$[\text{dex}]$&\\
\hline
\vspace{-0.2cm}$log(n_{\text{e}})$&\\\vspace{-0.2cm} &$3.4^{+0.1}_{-0.1}$&$3.4^{+0.1}_{-0.1}$\\
$[\text{cm}^{-3}]$&\\
\hline
\vspace{-0.2cm}$\text{Distance}$&\\\vspace{-0.2cm}&$2600^{+2500}_{-1300}$&$4400^{+3100}_{-2800}$\\
$[\text{pc}]$&\\
\hline
\vspace{-0.2cm}$\dot M$&\\\vspace{-0.2cm}&$110^{+120}_{-60}$&$30^{+80}_{-20}$\\
$[M_{\odot} \text{yr}^{-1}]$&\\
\hline
\vspace{-0.2cm}$\dot M v$&\\\vspace{-0.2cm}&$1.1^{+1.2}_{-0.6}$&$0.27^{+0.82}_{-0.18}$\\
$[10^{36} \text{ ergs cm}^{-1}]$&\\
\hline
\vspace{-0.2cm}$log({\dot E}_K)$&\\\vspace{-0.2cm}&$43.94^{+0.31}_{-0.32}$&$43.32^{+0.49}_{-0.61}$\\
$[\text{erg s}^{-1}]$&\\
\hline
\vspace{-0.2cm}${\dot E}_K/L_{Edd}$&\\\vspace{-0.2cm}&$0.099^{+0.12}_{-0.05}$&$0.023^{+0.053}_{-0.018}$\\
$[\text{\%}]$&\\
	\hline
	\end{tabular}
\end{table}
\section{Discussion}
\label{sec:discussion}
\subsection{AGN Feedback Contribution}
In order to be a major contributor to AGN feedback, S1 needs to have a kinetic energy of at least $\sim0.5\%$ \citep{2010MNRAS.401....7H} or perhaps as much as $\sim5\%$ \citep{2004ApJ...608...62S} of the quasar's Eddington luminosity ($L_{Edd}$), depending on the theoretical model. To find $L_{Edd}$, we follow the method by \citet{2022Byun}, finding the full-width half max (FWHM) of the \ion{Mg}{ii} emission in the SDSS spectrum, and using the \ion{Mg}{ii}-based equation by \citet{2019ApJ...875...50B} to find the mass of the black hole. As there is \ion{Fe}{ii} emission in the region of \ion{Mg}{ii} emission, we use the \ion{Fe}{ii} template by \citet{2006ApJ...650...57T} and run a best fit algorithm to match the template to the spectrum, following \citet{2018ApJ...859..138W}.

The resulting black hole mass is $M_{BH}=7.05^{+2.68}_{-2.03}\times10^{8} M_\odot$, with a corresponding Eddington luminosity of $L_{Edd}=8.89^{+3.38}_{-2.56}\times10^{46}$ erg s$^{-1}$. The Eddington ratio of the outflow ranges from $0.099^{+0.12}_{-0.05}\%$ (for solar metallicity) to $0.023^{+0.053}_{-0.018}\%$ (for super-solar metallicity), which is below the threshold for AGN feedback contribution.
\subsection{Comparison with Previous Work}
We have found the value of $n_H$ of S1 based on the ratios between the column densities of excited state and resonance state \ion{Fe}{ii}. This has notably done by \citet{2008ApJ...688..108K} for the outflow of the quasar NVSS J235953-124148. The $\log{n_e}$ values from the different energy states in Fig.~\ref{fig:ratioplot} are in agreement within $\sim0.2$ dex, which is comparable to the agreement shown in Fig. 3 of \citet{2008ApJ...688..108K}, showing that the ratios between \ion{Fe}{ii} energy states can be consistently used to probe the $n_e$ value and the distance of an outflow from its source.
\section{Summary and Conclusion}
\label{sec:conclusion}
We have identified two outflow systems from the VLT/UVES spectrum of the quasar SDSS J1439-0106, the mini-BAL S1 and the NAL S2. After measuring the column densities of the ions identified in S1, we used these measurements to find the $N_H$ and $U_H$ values of S1 via photoionization analysis, using models of both solar and super-solar metallicity (see Fig.~\ref{fig:nvuplots}).

With the abundance ratios between five different excited states of \ion{Fe}{ii} and the resonance state, we found the electron number density of S1 (see Fig.~\ref{fig:ratioplot}). We have also found its distance from the quasar, the mass flow rate, and kinetic luminosity. Based on the ratio between the kinetic luminosity of S1 and the Eddington luminosity of the quasar, we conclude that it is insufficient for the outflow system to contribute to AGN feedback.
\section*{Acknowledgements}
NA, DB and AW acknowledge support from NSF grant AST 2106249, and NASA STScI grants AR-15786, AR-16600, and AR-16601.

\section*{Data Availability}

The normalized UVES spectrum of J1439-0106 is part of the SQUAD database made available by \citet{michael_murphy_2018_1463251} and described by \citet{Murphy2019}. The SDSS spectrum is available in the SDSS database.



\bibliographystyle{mnras}
\bibliography{J1439} 





\bsp	
\label{lastpage}
\end{document}